\documentstyle[12pt,aaspp4]{article} 

\lefthead{CHIBA}
\righthead{DARK MATTER SUBSTRUCTURE}

\begin{document}

\title{Probing Dark Matter Substructure in Lens Galaxies}

\author{Masashi Chiba}
\affil{National Astronomical Observatory, Mitaka, Tokyo 181-8588, Japan\\
email: chibams@gala.mtk.nao.ac.jp}

\begin{abstract}
We investigate the effects of numerous dark matter subhalos in a galaxy-sized
halo on the events of strong lensing, to assess their presence as expected
from the cold dark matter scenario. Lens galaxies are represented by a smooth
ellipsoid in an external shear field and additional cold dark matter subhalos
taken from Monte Carlo realizations which accord with recent N-body results.
We also consider other possible perturbers, globular clusters and luminous
dwarf satellites, for comparison.
We then apply the models to the particular lens systems with four images,
B1422$+$231 and PG1115$+$080, for which smooth lens models are unable to
reproduce both the positions of the images and their radio flux ratios or
dust-free optical flux ratios simultaneously.
We show that the perturbations by both globular clusters and dwarf
satellites are too small to change the flux ratios, whereas cold dark matter
subhalos are most likely perturbers to reproduce the observed flux ratios in
a statistically significant manner. This result suggests us the presence
of numerous subhalos in lens galaxies, which is consistent
with the results of cosmological N-body simulations.
\end{abstract}


\section{Introduction}

The cold dark matter (CDM) scenario for structure formation in the universe
has been quite successful to explain a wide variety of observational results,
including characteristic fluctuations of the cosmic microwave background and
large-scale structure of galaxies on scales larger than $\sim 1$ Mpc.
The currently best world model, which accords with the results of
many cosmological observations, consists of approximately 30 \% CDM and
70 \% vacuum energy or quintessence (Bahcall et al. 1999).

However, the recent advent of high-resolution N-body simulations on
CDM-based structure formation has enabled to highlight various discrepancies
with existing observations on scales smaller than $\sim 1$ Mpc (see, e.g.,
Dav\'e et al. 2001 for each piece of conflicting evidence and relevant
references). One of the most serious issues is that CDM models predict
the existence of several hundred clumps or ``subhalos'' in
a galaxy-sized halo, in sharp contrast to the observed number of about
dozen Milky Way satellites (Klypin et al. 1999; Moore et al. 1999).
Also, this large number of subhalos appears to conflict with the observed
thinness of disk components, since they play a dynamical role in the heating
of disks (T\'oth \& Ostriker 1992). Recently proposed arguments to preclude
these difficulties in CDM models include: (1) formation of satellite galaxies
may have been largely suppressed by photoionization on a cosmological scale
(Bullock et al. 2000), and (2) CDM subhalos in the context of the vacuum-energy
dominated universe may not be efficient perturbers for the heating of disks
(Font \& Navarro 2001). If these arguments are the case, then we expect
the existence of many dark subhalos around a galaxy like the Milky Way.
Alternatively, CDM models may differ from their standard representation
on small scales (Spergel \& Steinhardt 2000; Kamionkowski \& Liddle 2000).

If CDM models on scales $\la 1$ Mpc remain valid in the above sense, then
the question arises as to {\it how we assess many dark subhalos in a galaxy?}
There is an intriguing possibility that some or all subhalos may associate
primordial gas which has failed to form stars, in the form of the observed
High Velocity Clouds (Blitz et al. 1999). However, observational evidence
is not fully convincing enough to accept this hypothesis.

Here, we focus on the effects of such mass substructure on strong lensing
phenomena, where a host galaxy or its halo plays a role in splitting an image
of a background source. Mao \& Schneider (1998, hereafter MS) pointed out
that in some lensed systems, any `simple' lens models, represented by a smooth
gravitational potential, fail to reproduce observed flux ratios
among multiple images, especially radio flux ratios which exempt from dust
extinction and microlensing in a lens galaxy, while image
positions are successfully recovered. They then considered the possibility
that flux ratios may be strongly affected by substructure in a lens, by
employing uniformly distributed point masses and plane-wave perturbations
to study the effects of globular clusters and spiral arms, respectively.
Applying their perturbed lens model to the four-image QSO, B1422$+$231,
they concluded that the required density perturbation for explaining
the observed discrepant flux ratios in this lensed QSO is of the order of
1 \% of the critical surface density in its lens galaxy.

Our aim in this paper is to investigate whether the existence of
numerous CDM subhalos in a galaxy is assessed by applying the MS methodology
to lens galaxies\footnote{Since this paper being submitted, a new
investigation has also appeared in astro-ph (Metcalf \& Madau 2001)
which explores similar issues and reaches compatible conclusions.}.
We consider three possible perturbers to lensed images:
globular clusters, luminous dwarf satellites, and CDM subhalos. In contrast
to MS, we take into account a realistic representation of globular clusters
by adopting their observed number and surface density distribution in
external galaxies (e.g. Harris 1991). The properties of CDM subhalos are
taken from published N-body simulation results (Klypin et al. 1999).
We consider two lens systems, B1422$+$231 (source redshift $z_S=3.62$,
lens redshift $z_L=0.34$) and PG1115$+$080 ($z_S=1.72$, $z_L=0.31$),
which exhibit most discrepant flux ratios among any other lens systems.
The foreground lens galaxies of these systems are known as elliptical
galaxies, so that the possibility of spiral-arm perturbations to the
lensed images is excluded. Unless otherwise stated,
we adopt the set of the cosmological parameters, $\Omega=0.3$, $\Lambda=0.7$,
and $h=0.7$ ($h \equiv H_0/100$ km~s$^{-1}$~Mpc$^{-1}$) in this work.

\section{Lens Model and Method}

\subsection{Smooth Lens Model}

B1422$+$231 shows the colinear, three highly magnified images A, B, and C,
and the additional faint image D located near the lens galaxy. This
configuration of the images suggests that the source is close to and inside
a cusp singularity provided by the lens. PG1115$+$080, also having four
lensed images, shows the close pair of the images A1 and A2, and this
configuration emerges if the source is close to and inside a fold caustic.
Figure 1 shows the configuration of the observed lensed images (open circles).
In such lens systems associated with a cusp or fold singularity, there
exist a universal relation between the image fluxes, i.e.,
(A+C)$/$B$=1$ or A2$/$A1$=1$, whereas the observed flux
ratios violate these rules significantly (MS; Impey et al. 1998),
as described below.
Several different lens models have been constructed to attempt to reproduce
both the image positions and flux ratios of these lens systems, but
all of the previous models have been unsuccessful (e.g., Hogg \& Blandford 1994;
Kormann et al. 1994; MS; Impey et al. 1998).
While each of the models differs in its parameterization, the observed
discrepancy in the flux ratios, compared with the expected universal relation
from a cusp or fold singularity, suggests that it is an intrinsic difficulty
for smooth lens models, not associated with a particular parameterization.

The choice of the smooth lens model is complicated by basically an infinite
number of degeneracies, where different models yield different image
amplifications. However, since the above issue of the failure to explain
the singularity-related image configurations and flux ratios are generic
in any smooth models, it may be sufficient to consider here a single
smooth lens model to be applied to both lens models, in order to clarify
to what characteristic extent each of the three perturbers, globular clusters,
dwarf satellites, and CDM subhalos, affects the concerned images.

Here, we select, as a smooth lens model, a singular isothermal ellipsoid (SIE)
in an external shear field (Kormann et al. 1994) to take advantage of its
simplicity. This model has been widely used in lens modeling and successful
to reproduce many other lens systems (e.g. Keeton et al. 1998).
An isothermal profile for the total mass distribution of elliptical galaxies is
well supported by the detailed dynamical studies of local ellipticals
(Gerhard et al. 2001), individual lens modeling, and statistics
(e.g. Maoz \& Rix 1993; Kochanek 1995; Grogin \& Narayan 1996).
The inclusion of an external shear field appears to be
necessary both to improve the fits of lens models to the data and
to make an axis ratio distribution of individual lenses being consistent with
the observed axis ratio distribution of light (Keeton et al. 1997).
We believe that this model provides a sufficient representation for
our purpose here to show the effects of substructure and also for comparison
with MS's work utilizing the same model. Our smooth model holds ten parameters:
critical angular scale
(Einstein radius in an angular scale) characterizing the strength of the
lens potential $\theta_0$, axis ratio $f$ of the lens, position angle
$\phi_g$ of its semi-minor axis, strength and direction of the external
shear $(\gamma,\phi_s)$, source position on the source plane, lens galaxy
position on the lens plane, and unlensed source flux.

The observational constraints are the positions and fluxes of lensed images,
and the galaxy position.
For B1422$+$231, we take {\it Hubble Space Telescope}
(HST) observations with the Faint Object Camera by Impey et al. (1996).
The positions and flux ratios with respect to the image B are taken from their
Table 1 and 2 (after correction of the typos noted by MS), and we adopt
the uncertainties, $0.^"01$ and 0.02, for image positions and their
flux ratios, respectively, and slightly more conservative uncertainty
for the galaxy position, $0.^"015$. It is worthwhile to note that the optical
flux ratio A$/$B ($=0.78\pm0.02$) deviates largely from the radio flux ratio
($=0.98\pm0.02$) (Patnaik et al. 1992), while
other flux ratios, C$/$B and D$/$B, are nearly the same in the optical and
radio. For PG1115$+$080, the image and galaxy
positions, flux ratios, and their uncertainties are taken from the observation
with the {\it HST} near-infrared camera multiobject spectrograph (NICMOS)
by Impey et al. (1998). PG1115$+$080 is radio quiet, but Impey et al. (1998)
concluded that the lens galaxy is dust free, as the flux ratios show
little variation with wavelength.

Given the above observational constraints, we undertake the $\chi^2$ fitting
to find the best model parameters. For B1422$+$231, we follow MS's procedure,
i.e. the discrepant flux ratio A$/$B between the optical and radio is excluded
in the $\chi^2$ measure. Figure 1 also shows the best-fit image positions
(crosses) and source position (solid circles), and Table 1
tabulates the list of some basic lens parameters, which are essentially the
same as and similar to the MS and Impey et al. results for B1422$+$231 and
PG1115$+$080, respectively. The model reproduces very
well the observed image positions within 1 $\sigma$ of the observational
uncertainties. However, notable discrepancies remain in the flux ratios,
as already noted by MS and Impey et al. (1998).

As mentioned above, the configuration of three bright images A, B, and C
in B1422$+$231 suggests that the source is close to and inside a cusp (as
shown in Figure 1), thereby leading to the expectation that the flux ratio
(A+C)$/$B is close to 1 (1.25 in the current model fitting). In contrast,
the radio observation shows the much larger flux ratio, $1.50\pm0.02$.
For PG1115$+$080, the close pair of A1 and A2 emerges if the images are
symmetrically arranged near a fold caustic (Figure 1) and the expected flux
ratio A2$/$A1 is unity (0.92 in the current fitting), whereas the
observed flux ratio, $0.65\pm0.02$, deviates significantly from 1.
These discrepant flux ratios have never been solved in previous lens
models with smooth mass distributions (MS; Impey et al. 1998).

\subsection{Substructure Models}

Given the above smooth lens potentials,
we consider three particular realizations for the substructure in lens
galaxies: globular clusters, luminous dwarf satellites, and CDM subhalos.
For all of these perturbing masses, we assume a point mass model, as their
spatial scales are negligibly small compared to dimensions of host galaxies
and their locations relative to undisturbed lensed images are generally
larger than their Einstein radii.

For globular clusters, we adopt the observed empirical relation
for their surface number density, $\Sigma_{gl}$, and total number, $N$,
in E/S0 galaxies (Harris 1976; 1991).
According to Harris, $\Sigma_{gl}$ is generally written as a power-law
function of the distance from the galactic center,
$\Sigma_{gl} \propto R^\alpha$, where $\alpha$ correlates with
absolute V magnitude of a host galaxy, $M_V$,
as $\alpha = -0.28 M_V - 7.9$. The coefficients in this relation correspond
to the case $h=0.75$.
The total number, $N$, can be derived from the relation between the
so-called specific frequency of clusters, $S_N$, and $M_V$ of a host galaxy,
as $S_N = N 10^{0.4 (M_V + 15)}$ (for $h=0.75$).
For the concerned lens ellipticals located in field or small groups,
{\it not clusters of galaxies}, the average value of $S_N$ is estimated
as $2.6\pm0.5$ (Harris 1991). The last two columns of Table 1 tabulate
our estimate of $\alpha$ and $N$, where we adopt $M_V=-20.62$ mag and $-20.86$
mag (for $h=0.75$) for the lens galaxies of B1422$+$231 and PG1115$+$080,
respectively, based on
the observed V magnitudes of lenses (Impey et al. 1996 for B1422$+$231;
Christian et al. 1987 for PG1115$+$080) and K corrections (Poggianti 1997).
We distribute these clusters with a fixed mass of $5\times 10^5$ $M_\odot$
randomly within a circle with angular radius $10 \theta_0$, which corresponds
to linear radii, $R_{max}$, of 38 $(h/0.7)^{-1}$ kpc and 52 $(h/0.7)^{-1}$ kpc
for the lens galaxies of B1422$+$231 and PG1115$+$080, respectively.

For dwarf satellites, we adopt their velocity distribution function
in the Local Group, as fitted by Klypin et al. (1999, their eq.~1):
$N(>V_c) = 385 (V_c/10{\rm km~s}^{-1})^{-1.3}$
($h^{-1}$~Mpc)$^{-3}$ for $R < 200$ h$^{-1}$ kpc and $V_c>10$ km s$^{-1}$,
where $V_c$ denotes the circular velocity. This function is then reduced
to the mass function, using the relation between $V_c$ and mass $M$
in the spherical collapse model (e.g. Navarro, Frenk, \& White 1997,
hereafter NFW),
$M = 2.2 \times 10^8 (V_c/10{\rm km~s}^{-1})^3$ $h^{-1}$~M$_\odot$
at redshift of the concerned lens galaxies of 0.3.
Thus, the cumulative mass function of satellites, $N(>M)$, within a spherical
volume with radius 200 $h^{-1}$~kpc is given by
\begin{equation}
N(>M) = 13 \left( \frac{M}{2.2\times 10^8 h^{-1}M_\odot} \right)^{-0.4} \ ,
\qquad {\rm for \ dwarf \ satellites} \ ,
\end{equation}
and the differential mass function has a form of
$dN/dM \propto M^{-1.4}$. The surface density distribution of satellites
is assumed to decline with radius as $R^{-0.8}$, as supported by
observations of various external galaxies (Lake \& Tremaine 1980;
Lorrimer et al. 1994). We assume $R_{max}=200$ $h^{-1}$ kpc
for the distribution of the dwarf galaxies, to be consistent with the adopted
velocity function.

For CDM subhalos, we adopt the result of the high-resolution,
cosmological N-body simulation by Klypin et al. (1999). Using their eq.(3)
for the velocity function of subhalos derived by the $\Lambda$CDM simulation
(i.e. for the case of $\Omega=0.3$, $\Lambda=0.7$, and $h=0.7$),
the cumulative mass function of subhalos within a spherical volume with
radius 200 $h^{-1}$~kpc is given by
\begin{equation}
N(>M) = 168 \left( \frac{M}{2.2\times 10^8 h^{-1}M_\odot} \right)^{-0.9} \ ,
\qquad {\rm for \ CDM \ subhalos} \ ,
\end{equation}
and $dN/dM \propto M^{-1.9}$. Klypin et al.'s result for the $\Lambda$CDM
simulation is formally valid for $V_c > 20$ km s$^{-1}$, but we extrapolate
their result to $V_c = 10$ km s$^{-1}$ (or $M = 2.2\times 10^8$
$h^{-1}M_\odot$) as their paper argues based on the higher
resolution simulation.
The surface density distribution of these subhalos is derived from the
projection of the NFW profile, since their volume density appears to be
well represented by the NFW profile (Ghigna et al. 2000).
We adopt the length scale of the NFW profile as $r_s = 20$ $h^{-1}$ kpc
and $R_{max}=200$ $h^{-1}$ kpc.

Figure 2 shows the adopted cumulative mass functions of dwarf satellites
(dashed line) and CDM subhalos (solid line) inside $R=200$ $h^{-1}$~kpc.
As was emphasized by Klypin et al. (1999) and Moore et al. (1999),
the number of CDM subhalos with mass greater than
$10^8-10^9$ $M_\odot$ is an order of magnitude higher than that of
the observed dwarf satellites in the Local Group.

The presence of point masses around a host galaxy as described above
yields a perturbed lensing potential,
\begin{equation}
\delta\psi = \sum_{i=1}^N \frac{\tilde{m}_i}{2}
             \ln [(x_1-x_{1i})^2+(x_2-x_{2i})^2]  \ ,
\end{equation}
where
$\tilde{m}_i \equiv M_i / (\Sigma_{cr}\pi D_{d}^2\theta_0^2) $.
$(x_1,x_2)$ and $(x_{1i},x_{2i})$, $i=1 ... N$, denote the positions of
an unperturbed lensed image and $N$ point masses (having mass $M_i$)
on the lens plane, respectively. The critical surface density $\Sigma_{cr}$
is defined as $\Sigma_{cr} = c^2/(4\pi G) \times D_d D_s / D_{ds}$, where
$D_d$, $D_s$, and $D_{ds}$ are angular diameter distances to the lens, to the
source, and between the lens and the source, respectively.
We then generate these point masses on the lens plane by Monte Carlo
simulations following the spatial and mass distributions described above,
calculate the shear caused by the point-mass lenses
$(\delta \gamma_1, \delta \gamma_2)$ (where perturbing convergence is zero),
and estimate the total magnification of each lensed image, as was explored
by MS.

\section{Results}

\subsection{Perturbed Flux Ratios}

We plot, in upper panels of Figure 3, the probability distributions of
amplification ratios [in units of flux B in B1422$+$231 (left panel)
and flux A1 in PG1115$+$080 (right panel)],
for globular clusters (dashed line) and dwarf satellites (solid line)
as perturbers. Solid bars denote the observed flux ratios of A$/$B
and $r$ (see below) in B1422$+$231 and A2$/$A1 in PG1115$+$080,
respectively. As is evident, both perturbers are not sufficient
to reproduce the observed flux ratios; the probability distributions
are essentially represented by $\delta$ functions. Left panel
for B1422$+$231 also shows  the probability distribution of a flux ratio,
$r \equiv (A+B+C)/(|A|+|B|+|C|)$ introduced by MS.
If these three images obey the prediction of a cusp singularity, then $r=0$.
Our smooth lens model yields $r=0.113$, whereas the observed radio fluxes
give $r=0.2$. It is clear that the probability of $r$ having the value
larger than the observed one (0.2) is null for globular clusters, whereas
it is only 3.4 \% for dwarf galaxies. For PG1115$+$080, the probability
of A2$/$A1 having a deviation smaller than the observed one (0.65) is null
for globular clusters and 9.2 \% for dwarf satellites.

On the other hand, as is shown in lower panels of Figure 3, the presence
of CDM subhalos yields the much larger probability of recovering the
observed flux ratios: the probability of A$/$B having a deviation larger
than the observed one (0.98) is 21.3 \% (25.3 \% for $r$ having a deviation
larger than 0.2) for B1422$+$231, and the probability of A2$/$A1
having a deviation smaller than the observed one is 15.4 \% for PG1115$+$080.
Thus, our present experiments suggest that numerous CDM subhalos around the
concerned lens galaxies explain more likely the observed flux ratios
of the lensed images than globular clusters and dwarf galaxies.
It is interesting to note that the probability distribution is more
broadly distributed for higher magnification images, which is consistent
with MS's analytical result. This implies that the effects of CDM halos
are more easily seen in the very bright images, which are associated
with a fold or cusp singularity, than other ordinary lensed images.

We have also examined the most realistic model by considering the presence
of all three populations in the lens mass distribution. For B1422$+$231,
the resultant distributions of the amplification ratios remain essentially
the same as those for CDM subhalos only, whereas for PG1115$+$080,
the probability of A2$/$A1 having a deviation smaller than the observed one
is somewhat boosted, about 5 \% larger than the case of CDM subhalos only.
It also appears that neither of the two lens systems are affected by
the presence of globular clusters.

So far, our estimation of flux ratios implicitly assumes that the additional
amplification of each image provided by substructure perturbations occurs
independently of the events for other images. However, there is a possibility
that more than one image are simultaneously affected by the same perturbations:
if an ensemble of subhalos around the observed images, {\it not just one
subhalo located close to only one of images}, are involved in the
required perturbations to flux ratios, then there may exist a finite
correlation between the amplification factors of images.

Figure 4 shows, in the case of CDM subhalos, the amplification factors
of two specific images in B1422$+$231 (panels a and b) and PG1115$+$080
(c and d). The dotted lines denote the unperturbed amplification factors
and solid lines indicate the observed flux ratios. It follows that in
B1422$+$231, the perturbations to the images B and A are somewhat correlated,
whereas those to the images B and C are uncorrelated. Also in PG1115$+$080,
the perturbations to the images A1 and A2 are somewhat correlated
but those to the images A1 and B are uncorrelated. These properties suggest
that the reproduction of the observed flux ratios is associated with
an ensemble of subhalos around the images B and A in B1422$+$231 and around
the images A1 and A2 in PG1115$+$080. We have actually identified some
number of subhalos around the above correlated images, by plotting
the positions of subhalos when the observed flux ratios are reproduced.

This somewhat local nature of the perturbations is also the case for dwarf
satellites, but the significance of the correlation between the amplification
factors of images turns out to be slightly reduced. As a measure of
correlation, we estimate the linear correlation coefficient, $r_1$,
between the images B and A in B1422$+$231 and
between the images A1 and A2 in PG1115$+$080.
For B1422$+$231, we obtain $r_1=1.94 \times 10^{-3}$ and
$-5.72 \times 10^{-3}$ for dwarf satellites and CDM subhalos, respectively,
whereas for PG1115$+$080, $r_1=-6.40 \times 10^{-4}$ and $-7.71 \times 10^{-3}$,
respectively. Although the significance of these difference in $r_1$ between
dwarf satellites and CDM subhalos is not high (with the probability of 0.86,
where a lower probability indicates a higher significance), we have confirmed,
by plotting the positions of dwarf satellites, that {\it only one} satellite,
located in the vicinity of either the image B or A in B1422$+$231
and either A1 or A2 in PG1115$+$080, plays a role in reproducing
the observed flux ratios.

\subsection{Perturbed Deflection Angles}

We further calculate the perturbed deflection angles of the four macro images,
$\delta\alpha$, caused by the perturbers. Figure 5a shows, for B1422$+$231,
the probability distributions $p(\delta\alpha)$ of $\delta\alpha$ for the four
images, caused by globular clusters. The deflection of the images A, B, and C
amounts to only of the order of 0.$^"$001, whereas the image D is subject to
a systematically larger deflection, $\sim 0.^"0035$, possibly due to
its position very close to the center of the lens galaxy, where larger number
of clusters are distributed than at the locations of other images.

Figure 5b shows the cases of dwarf satellites (dotted lines)
and CDM subhalos (solid lines) depending on the range of the flux ratio $r$.
In upper panel, $r$ is close to the value obtained from the undisturbed, smooth
lens model ($0.09\le r \le 0.13$), i.e. near the peak of its probability
distribution shown in Figure 3, whereas lower panel corresponds to the range
near the observed value ($0.18\le r \le 0.22$).
It follows that dwarf satellites produce the deflection angles less than
0.$^"$01 with a peak at $\sim 0.^"003$ for $0.09\le r \le 0.13$, whereas
the reproduction of the observed flux ratio is accompanied by much larger
deflection angles, as is seen for $0.18\le r \le 0.22$. This is consistent
with the fact that the reproduction of the observed flux ratios is associated
with one dwarf satellite in the vicinity of one of the images, so that
the deflection angle is large.

Figure 5b also indicates that CDM subhalos yield a systematically larger
deflection than other perturbers: a typical value of $\delta\alpha$ ranges
from 0.$^"$01 to 0.$^"$02. Also, $p(\delta\alpha)$ in the case of
$0.18\le r \le 0.22$ including the observed value $r=0.20$ is basically
similar to the case of $0.09\le r \le 0.13$, thereby indicating
that the observed flux ratio which largely deviates from the smooth-lens
prediction is {\it not} accompanied by very large deflection of one of the
images; an additional amplification of the images is supplied by an ensemble
of CDM subhalos around more than one image,
not just one subhalo being very close to one of images.

\section{Discussion and Concluding Remarks}

We have investigated the possibility that the large number of CDM subhalos
in a galaxy-sized halo, as suggested by recent high-resolution N-body
simulations, can affect flux ratios in multiply-imaged lens systems
in a statistically realistic level. Our Monte Carlo models based on the
N-body results have shown that the observed flux ratios in B1422$+$231
and PG1115$+$080, both of which deviate significantly from the predictions of
any simple lens models, can be reproduced by the perturbations of CDM subhalos,
with a finite probability ranging from 15 \% to 25 \%. The corresponding
perturbation to a smooth lens potential is of the order of 1 \% of the critical
surface density, which is consistent with the MS result based on their toy
model. On the contrary, the realizations of globular clusters, in accordance
with the observed properties of their specific frequency and spatial
distribution, are unable to produce any change in flux ratios, whereas
luminous dwarf satellites as observed in the Local Group are unsatisfactory
to explain the observed flux ratios in a statistically significant level.

Our experiments have also indicated that lensed images with higher
magnifications are more easily affected by substructure in a lens, which
is consistent with MS's analytic result. In this sense, the lens systems
employed in this work, B1422$+$231 and PG1115$+$080, are particular examples,
because of the large magnifications of the images A, B and C in B1422$+$231
associated with a cusp singularity, and of the images A1 and A2 in PG1115$+$080
associated with a fold singularity. Interestingly, quite a similar lens
configuration to PG1115$+$080, MG0414$+$0534, having four images, holds a
close pair of bright images A1 and A2 and the flux ratio A1$/$A2 ranges from
$\sim 2$ in $I$ band to $\sim 1.1$ at 8 GHz (Falco, Leh\'ar, \& Shapiro 1997).
As the radio flux ratio is close to unity as predicted from a smooth lens
model with a fold singularity, this system may be affected by either
of microlensing by stellar mass objects (Witt, Mao, \& Schneider 1995) or
dust extinction in the lens (Angonin-Willaime et al. 1999). Searches for
other similar lens systems to PG1115$+$080 will be very useful for setting
limits on the probability of showing a discrepant flux ratio. Besides the
above particular cases, we anticipate that the effects of CDM subhalos
on lensed images are less significant for other lens systems without
association of a cusp or fold singularity, and the effects of dwarf satellites
may be essentially null for such lens systems, although it is beyond the
scope of the paper to consider it in more detail.

As mentioned above, microlensing events by stellar mass objects, which also
affect brightness of background sources, are unlikely to explain the radio flux
ratios of the concerned lens systems, as the radio sources have a much larger
extension than the Einstein radius of a stellar mass. Also, if the luminosity
profile of the lens galaxies is represented by a de~Vaucouleurs law, the
effective radius is only 0.$^"$6 for the lens of PG1115$+$080 (Impey et al.
1998), compared to 1.$^"$1 -- 1.$^"$2 for the positions of the images A1 and A2
from the lens center, and the lens of B1422$+$231 has probably a similar
effective radius (because of similarity in its magnitude and redshift to
the former case), compared to the positions of the images A, B and C (1.$^"$0
-- 1.$^"$1 from the center). Thus, the probability of microlensing by
stars is expected to be small for the concerned images.

We finally remark that in the highly magnified lens systems as considered
here, one should cautiously use the constraints on lens parameters obtained
from radio fluxes, because they are easily affected by CDM subhalos.
The perturbation effect of substructure on the lensing is boosted by
the undisturbed amplification factor of images (Metcalf \& Madau 2001),
so that the highly magnified images associated with a singularity yield
only weak constraints on the model; other, more weakly amplified images
provide stronger constraints (MS). Also, as was already mentioned in MS,
the typical perturbations of CDM subhalos to lens potentials, of the order
of only 1 \%, are too small to affect the determination of $H_0$ from the
time delay measurement.

\acknowledgments
The author is grateful to N. Sugiyama, T. Hamana and K. Otsuki for their
useful comments on this work. He also thanks the anonymous referee for
useful suggestions and comments on the manuscript.

\clearpage


\begin{deluxetable}{lcccccccccc}
\footnotesize
\tablenum{1}
\tablewidth{0pc}
\tablecaption{Lens Parameters}
\tablehead{
\colhead{QSO}                & \colhead{$\theta_0$}         &
\colhead{$f$}                & \colhead{$\phi_g$}           &
\colhead{$\gamma$}           & \colhead{$\phi_s$}           &
\colhead{$\mu_i$($i=1-4$)\tablenotemark{a}}
                      & \colhead{$\alpha$\tablenotemark{b}} &
\colhead{$N$\tablenotemark{b}} \\
\colhead{}              & \colhead{('')}       &
\colhead{}              & \colhead{($^\circ$)} &
\colhead{}              & \colhead{($^\circ$)} &
\colhead{}              & \colhead{}           &
\colhead{} }
 
\startdata
B1422$+$231   & 0.78 & 0.84 & $-$57.5 & 0.21 & $-$53.9 &
   A:7.54, B:$-$9.94, C:4.93, D:$-$0.38 &
   $-$2.13 & 460 \nl
PG1115$+$080  & 1.15 & 0.86 & 102.2 & 0.11 & 52.5 &
   A1:13.53, A2:$-$12.38, B:$-$3.03, C:3.79 &
   $-$2.06 & 574 \nl
\enddata
\tablenotetext{a}{Magnification factors for four lensed images.}
\tablenotetext{b}{Power-law index $\alpha$ for the surface distribution of
globular clusters and their total number $N$, based on the estimate of
$M_V=-20.62$ mag and $-20.86$ mag (for $h=0.75$) for the lens galaxies of
B1422$+$231 and PG1115$+$080, respectively.}
\end{deluxetable}

\clearpage

\begin{figure}
\plotone{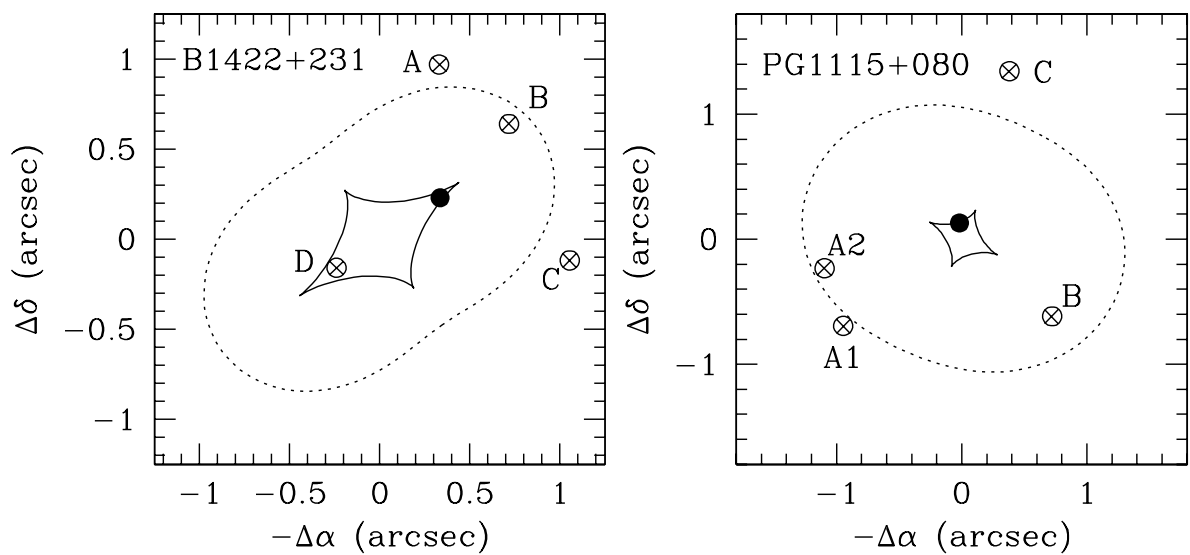}
\caption{
The lens configurations of B1422$+$231 and PG1115$+$080. Open circles
show the observed positions of the images, whereas crosses show the best-fit
positions of the images in our lens models. The source positions are indicated
with solid circles. Solid and dotted lines denote the caustics and critical
curves. Note that these lens systems are characterized by a cusp and fold
singularities in the positions of the sources, respectively.}
\end{figure}

\clearpage

\begin{figure}
\plotone{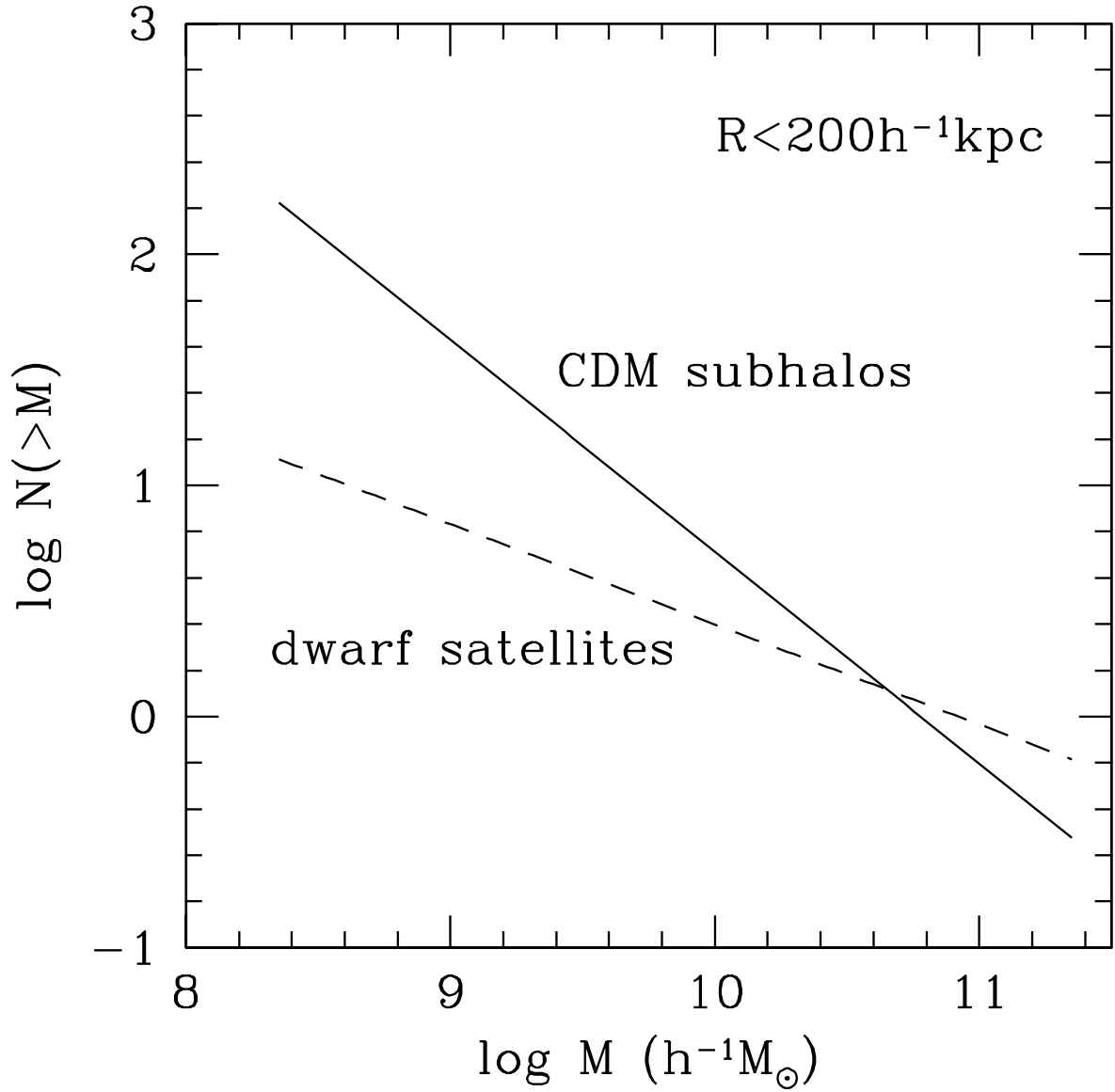}
\caption{
Adopted cumulative mass functions of dwarf satellites (dashed line)
and CDM subhalos (solid line) inside $R=200$ $h^{-1}$~kpc.}
\end{figure}

\clearpage

\begin{figure}
\plotone{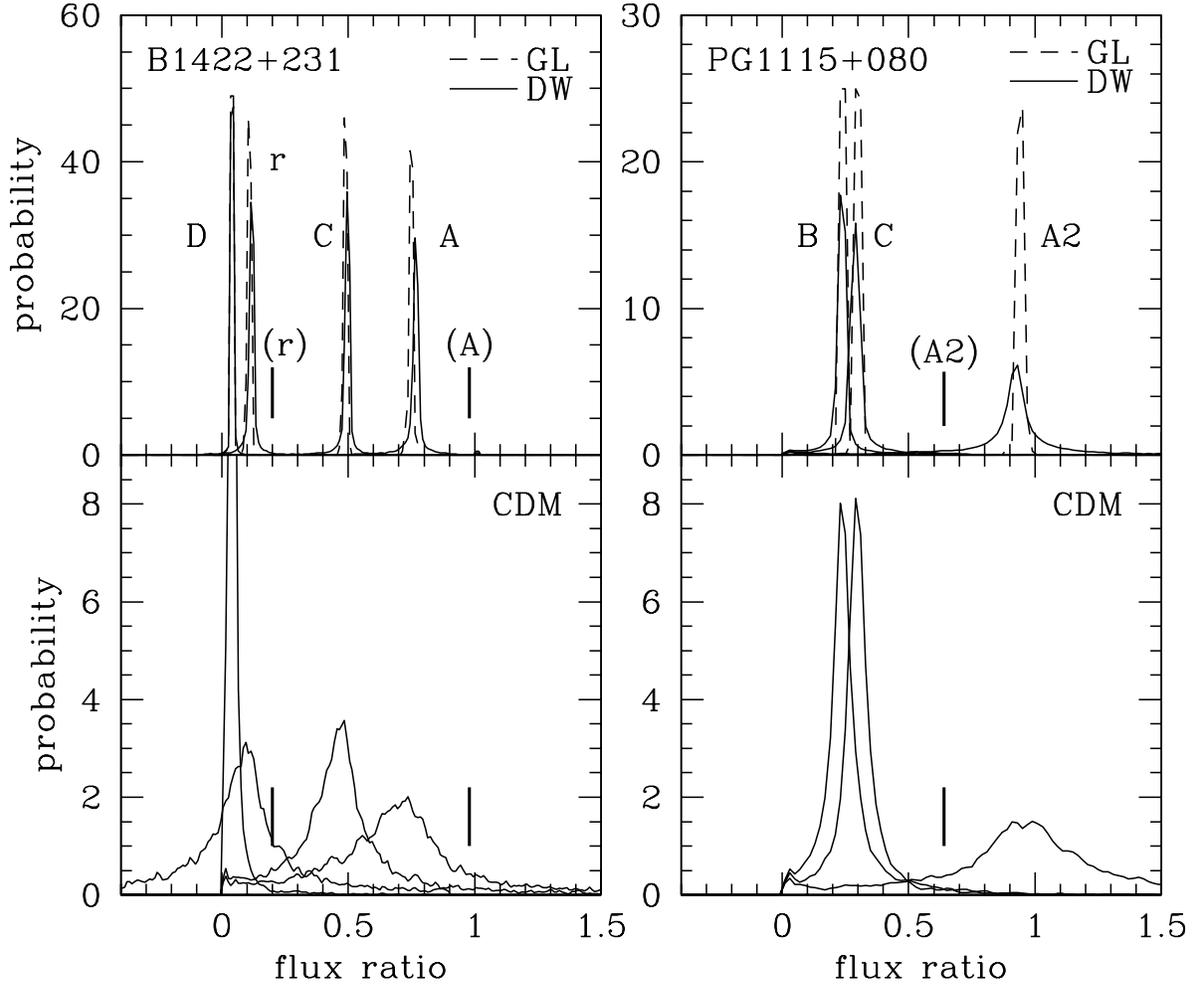}
\caption{
The probability distributions of the flux ratios for B1422$+$231 (left panels,
in units of flux B) and for PG1115$+$080 (right panels, in units of flux A1).
Upper panels show the cases of globular clusters (dashed lines) and
dwarf satellites (solid lines) as perturbers, whereas lower panels show the
case of CDM subhalos. Solid bars denote the observed flux ratios of A$/$B
and $r \equiv (A+B+C)/(|A|+|B|+|C|)$ for B1422$+$231 and A2$/$A1 for
PG1115$+$080.}
\end{figure}

\clearpage

\begin{figure}
\plotone{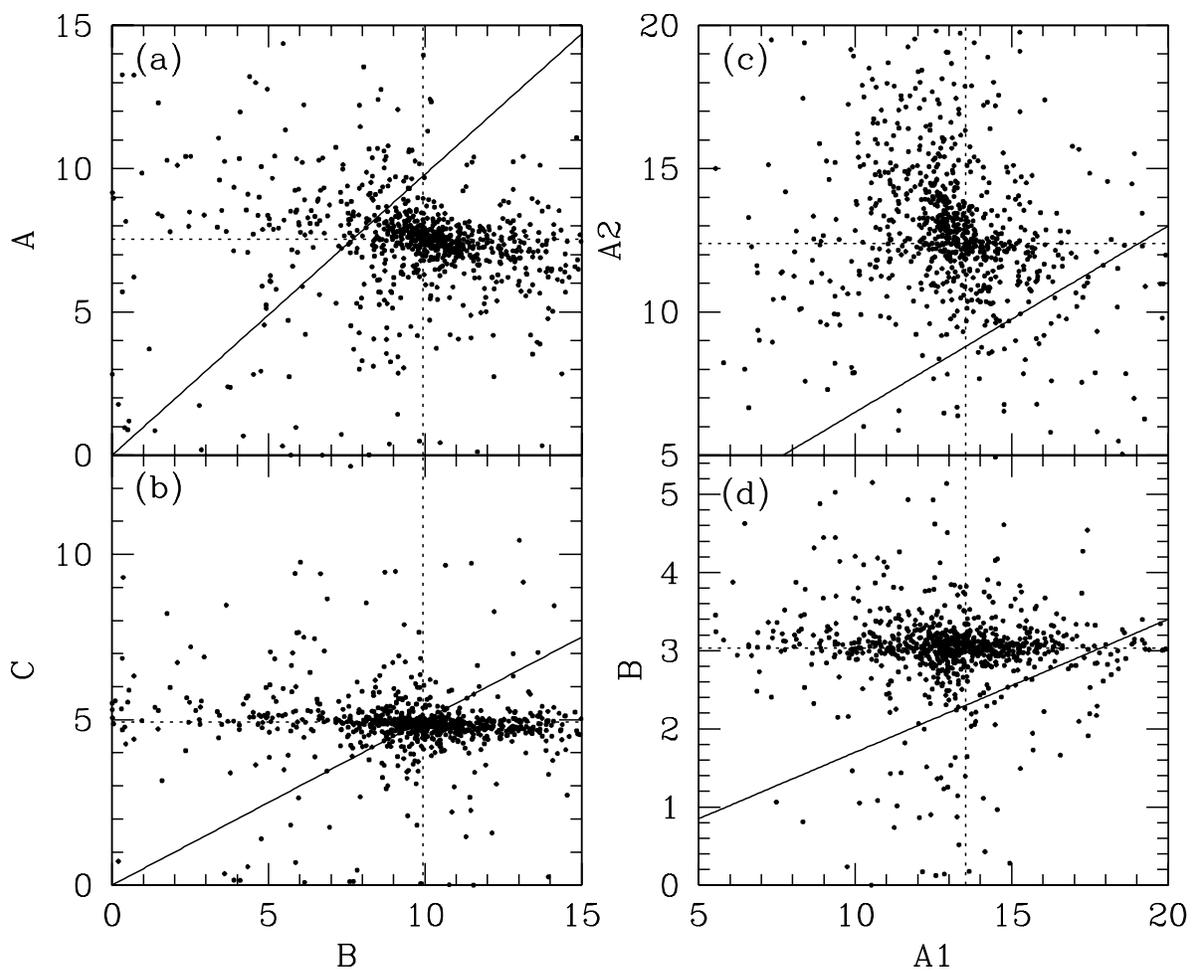}
\caption{
Comparisons between amplification factors of two specific images
in B1422$+$231 (panels a: B vs A, panel b: B vs C) and in PG1115$+$080
(panel c: A1 vs A2, panel d: A1 vs B), in the case of CDM subhalos.
Dotted lines denote the unperturbed amplification factors and solid lines
indicate the observed flux ratios.}
\end{figure}

\clearpage

\begin{figure}
\plotone{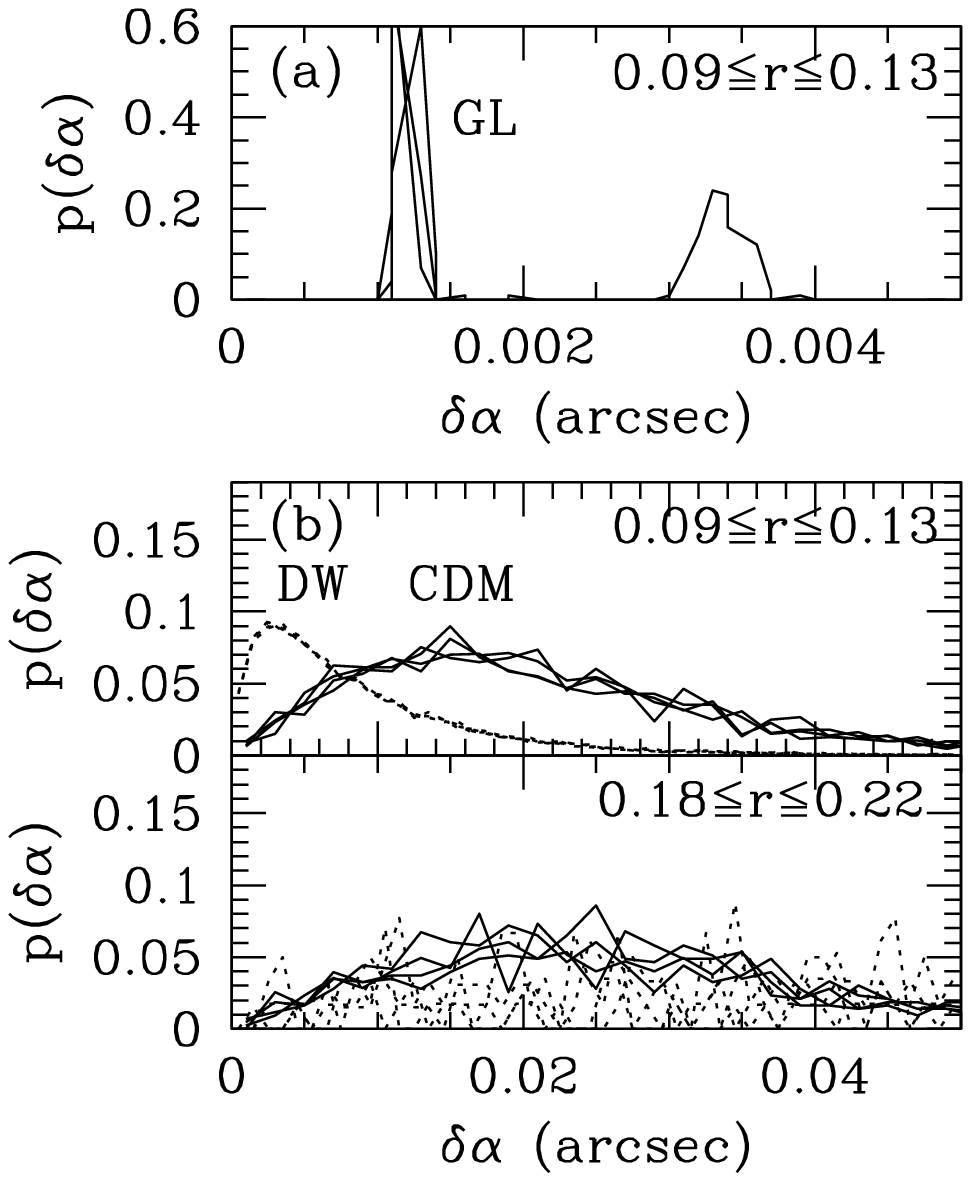}
\caption{
The probability distributions $p(\delta\alpha)$ of the perturbed deflection
angles $\delta\alpha$ for the four images of B1422$+$231, caused by globular
clusters [panel (a), solid lines], dwarf satellites [panel (b),
dotted lines)], and CDM subhalos [panel (b), solid lines].
In panel (b), upper panel shows the case of $0.09\le r \le 0.13$, i.e.
$r$ is close to the value from the prediction of the undisturbed, smooth lens
model ($r=0.113$), whereas lower panel corresponds to the range
$0.18\le r \le 0.22$ which includes the observed value ($r=0.20$).}
\end{figure}


\clearpage

\begin{thebibliography}{}

\bibitem[]{} Angonin-Willaime,~M.-C., Vanderriest,~C., Courbin,~F.,
Burud,~I., Magain,~P., \& Rigaut,~F. 1999, \aap, 347, 434

\bibitem[]{} Bahcall,~N.~A., Ostriker,~J.~P., Perlmutter,~S.,
\& Steinhardt,~P.~J. 1999, Science, 284, 1481

\bibitem[]{} Blitz,~L., Spergel,~D.~N., Teuben,~P.~J., Hartmann,~D.,
\& Burton,~W.~B. 1999, \apj, 514, 818

\bibitem[]{} Bullock,~J.~S., Kravtsov,~A.~V., \& Weinberg,~D.~H. 2000,
\apj, 539, 517

\bibitem[]{} Christian,~C.~A., Crabtree,~D., \& Waddell,~P. 1987, \apj,
312, 45

\bibitem[]{} Dav\'e,~R., Spergel,~D.~N., Steinhardt,~P.~J., \& Wandelt,~B.~D.
2001, \apj, 547, 574

\bibitem[]{} Falco,~E.~E., Leh\'ar,~J., \& Shapiro,~I.~I. 1997, \aj, 113, 540

\bibitem[]{} Font,~A.~S., \& Navarro,~J.~F. 2001, submitted to \apj \ Letters
(astro-ph/0106268)

\bibitem[]{} Gerhard,~O., Kronawitter,~A., Saglia,~R.~P., \& Bender,~R.
2001, \aj, 121, 1936

\bibitem[]{} Ghigna,~S., Moore,~B., Governato,~F., Lake,~G., Quinn,~T.,
\& Stadel,~J. 2000, \apj, 544, 616

\bibitem[]{} Grogin,~N.~A., \& Narayan,~R. 1996, \apj, 464, 92
(erratum 473, 570)

\bibitem[]{} Harris,~W.~E. 1976, \aj, 91, 822

\bibitem[]{} Harris,~W.~E. 1991, \araa, 29, 543

\bibitem[]{} Hogg,~D.~W., \& Blandford,~R.~D. 1994, \mnras, 268, 889

\bibitem[]{} Impey,~C.~D., Foltz,~C.~B., Petry,~C.~E., Browne,~I.~W.~A.,
\& Patnaik,~A.~R. 1996, \apj, 462, L53

\bibitem[]{} Impey,~C.~D., Falco,~E.~E., Kochanek,~C.~S., Leh\'ar,~J., \&
McLeod,~B.~A., Rix,~H.-W., Peng,~C.~Y., \& Keeton,~C.~R. 1998, \apj, 509, 551

\bibitem[]{} Kamionkowski,~M., \& Liddle,~A.~R. 2000, \prl, 84, 4525.

\bibitem[]{} Keeton,~C.~R., Kochanek,~C.~S., \& Seljak,~U. 1997, \apj,
482, 604

\bibitem[]{} Keeton,~C.~R., Kochanek,~C.~S., \& Falco,~E.~E. 1998, \apj,
509, 561

\bibitem[]{} Klypin,~A., Kravtsov,~A.~V., Valenzuela,~O., \& Prada,~F.
1999, \apj, 522, 82

\bibitem[]{} Kochanek,~C.~S. 1995, \apj, 445, 559

\bibitem[]{} Kormann,~R., Schneider,~P., \& Bartelmann,~M. 1994,
\aap, 284, 285

\bibitem[]{} Lake,~G., \& Tremaine,~S. 1980, \apj, 238, L13

\bibitem[]{} Lorrimer,~S.~J., Frenk,~C.~S., Smith,~R.~M., White,~S.~D.~M.,
\& Zaritsky,~D. 1994, \mnras, 269, 696

\bibitem[]{} Mao,~S., \& Schneider,~P. 1998, \mnras, 295, 587 (MS)

\bibitem[]{} Maoz,~D., \& Rix,~H.-W. 1993, \apj, 416, 425

\bibitem[]{} Mateo,~M.~L. 1998, \araa, 36, 435

\bibitem[]{} Metcalf,~R.~B., \& Madau,~P. 2001, \apj, in press
(astro-ph/0108224)

\bibitem[]{} Moore,~B., Ghigna,~S., Governato,~F., Lake,~G., Quinn,~T.,
\& Stadel,~J. 1999, \apj, 524, L19

\bibitem[]{} Navarro,~J.~F., Frenk,~C.~S., \& White,~S.~D.~M. 1995,
\mnras, 275, 56

\bibitem[]{} Patnaik,~A.~R., Browne,~I.~W.~A., Walsh,~D., Chaffee,~F.~H.,
\& Foltz,~C.~B. 1992, \mnras, 259, 1p

\bibitem[]{} Poggianti,~B.~M. 1997, \aaps, 122, 399

\bibitem[]{} Remy,~M., Surdej,~J., Smette,~A., \& Claeskens,~J.-F. 1993,
\aap, 278, L19

\bibitem[]{} Spergel,~D.~N., \& Steinhardt,~P.~J. 2000, Phys. Rev. Lett.,
84, 3760

\bibitem[]{} T\'oth,~G., \& Ostriker,~J.~P. 1992, \apj, 389, 5

\bibitem[]{} Witt,~H.~J., Mao,~S., \& Schneider,~P. 1995, \apj, 443, 18

\end{thebibliography}
\end{document}